# 2.4 GHz Flip-flop Device within Nonequilibrium Superconducting Diode

Xiangyu Bi[1,2†], Hongyi Li[1,2†], Aoshen Yang[1,2†], Yuqiang Fang[3†], Ganyu Chen[1,2], Shichong Yang[1,2], Yicheng Shen[1,2], Qizheng Sun[1,2], Junwei Huang[1,2*], Wei Jiang[1,2], Da Wang[1,4*], Fuqiang Huang[3,5*], Haijun Zhang[1,4], Qianghua Wang[1,4*], Hongtao Yuan[1,2*]

[1] *National Laboratory of Solid State Microstructures and Collaborative Innovation Center of Advanced Microstructures, Nanjing University, Nanjing 210093, China.*

[2] *College of Engineering and Applied Sciences, Nanjing University, Nanjing 210023, China.*

[3] *School of Materials Science and Engineering, Zhangjiang Institute for Advanced Study, Shanghai Jiao Tong University, Shanghai 200240, China.*

[4] *School of Physics, Nanjing University, Nanjing 210093, China.*

[5] *State Key Laboratory of High Performance Ceramics and Superfine Microstructure, Shanghai Institute of Ceramics, Chinese Academy of Science, Shanghai, 200050, China.*

[†]These authors contributed equally to this work.

*Correspondence to: htyuan@nju.edu.cn (H.T.Y.); qhwang@nju.edu.cn (Q.H.W.); huangfq@mail.sic.ac.cn (F.Q.H.); dawang@nju.edu.cn (D.W.); junweihuang@nju.edu.cn (J.W.H.).

**Superconducting diode effect exhibits asymmetric critical supercurrent and has profound implications for condensed matter physics. The technical appeals of such superconducting diodes are their ultrahigh on-off ratio and diode efficiency for superconducting electronics owing to the dissipationless supercurrent therein. However, realizing superconducting diode operation at high working frequency, which is a key requirement for practical applications, remains elusive and challenging. Here, we demonstrate a polarity-controllable superconducting diode with non-equilibrium Josephson junction and its edge-triggered flip-flop operation at a high frequency up to 2.4 GHz, within a van der Waals superconductor 2M-$WS_2$. By simply tuning the thickness of superconducting 2M-$WS_2$ nanoflakes to engineer inversion asymmetry in the junction, we achieve a high diode efficiency of 67% and an on-off ratio exceeding $10^5$. Importantly, the pulse width and duty cycle of output pulse signals in such superconducting diode flip-flop devices can be controlled in a broadband frequency range crossing 12 orders of magnitude. Theoretical analysis reveals that the non-equilibrium dynamic nature of supercurrent in these Josephson junctions enables such a high diode operating frequency and the polarity control of supercurrent. The 2.4 GHz non-equilibrium Josephson diode developed here provides a promising platform for advanced superconducting logic circuits and broadband telecommunication applications.**

Superconducting diode effect manifests nonreciprocal electronic transport in superconductors or Josephson junctions, playing significant roles in understanding emergent symmetry breaking phenomena in condensed matter physics[1]. In principle, breaking time-reversal symmetry, for example, applying an external magnetic field[2-5] or designing special heterostructures[6,7] is required for realizing superconducting diode effect in thermodynamic equilibrium states[8-15]. In contrast, in those non-equilibrium conditions, e.g., for Josephson junctions under an external electrical voltage[16], the strict requirement of time-reversal symmetry breaking[17] can be relaxed, since non-equilibrium charge dynamics caused by electrical current noise can trigger the tunneling among various metastable electronic states[16] and may further result in the superconducting diode effect. Considering the inversion symmetry breaking, the charge accumulation in an asymmetrically-stacked junction can be electronically asymmetric with respect to the applied voltage across the junction, which can lead to the diode effect and be well-described by the nonlinear quantum capacitance mechanism[18,19]. Notably, such a theory provides a strategy to realize superconducting diode effect without the well-known strict requirements of breaking time-reversal symmetry, and suggests an ultrahigh Terahertz-scale Josephson frequency within the order of $2eV/h \sim 10^{12}$ Hz ($h$ is Planck's constant, $e$ is the elementary charge, and assume electrical voltage $V \sim 1$ mV), which predicts a promising approach to realizing ultrahigh working frequency and broadband diode operation. However, experimentally achieving such a theoretically-proposed high frequency superconducting diode remains elusive.

In this Article, we demonstrate the 2.4 GHz edge-triggered flip-flop superconducting diodes based on non-equilibrium Josephson junction formed with a van der Waals superconductor 2M-$WS_2$. Unlike the 2H-$NbSe_2$ homojunction where the superconducting diode effect only exists by applying external magnetic fields, the 2M-$WS_2$/2M-$WS_2$ superconducting diodes can exhibit field-free diode effect with a high on-off ratio ($10^5$) of output voltage and a large diode efficiency $\eta$ (67%), and provide a unique platform to study nonreciprocal transport phenomena without the necessity to break time-reversal symmetry. We found that our designed edge-triggered flip-flop devices based on such 2M-$WS_2$ superconducting diodes maintain the capability of rectification and can generate stable and designable pulse signal at a high frequency up to 2.4 GHz, directly bridging the gap between theoretical predictions and practical high-frequency operation. More importantly, the polarity of these superconducting diodes can be reliably controlled simply by changing the thickness of superconducting nanoflakes and resulting inversion asymmetry of the homojunction. Based on our non-equilibrium diode-rectification simulations with the Resistively and Capacitively Shunted Junction (RCSJ) model, the amplitude of electronic circuit noise caused by thermal fluctuation in such asymmetric Josephson junctions plays the key role in inducing the superconducting diode effect without breaking time-reversal symmetry, and a high working frequency even up to sub-THz level can be expected theoretically based on such a model. Our results on high-frequency flip-flop operation within non-equilibrium superconducting diode pave the way toward superconducting logic circuits and next-generation telecommunication.

**Non-equilibrium field-free superconducting diode**

We chose the van der Waals superconductor 2M-$WS_2$ as the target material for investigating superconducting diode effect because of the following three reasons. First, owing to the unique thickness-dependent superconductivity of 2M-$WS_2$, the thin 2M-$WS_2$ nanoflakes below 10 nm can demonstrate the topological superconductivity while the thick 2M-$WS_2$ samples normally behave as conventional *s*-wave superconductors[20-23]. Thus, the 2M-$WS_2$ homojunction stacked of two nanoflakes with different thickness can naturally break the inversion symmetry and can induce charge accumulation asymmetry when applying an external voltage across the junction (Fig. 1a,b, and theoretical analysis on nonreciprocal supercurrent shown in Supplementary Figs. 1, 2), serving as the key aspect to generate the magnetic-field-free superconducting diode effect with a large nonlinear quantum capacitance (experimental details in Supplementary Figs. 3, 4). Second, the reported Majorana bound state of 2M-$WS_2$ nanoflakes[24,25] can theoretically induce asymmetric behavior in phase-dependent Andreev spectra owing to the applied Zeeman fields along the spin-orbit axis, leading to nonreciprocal supercurrents in the Josephson junctions and thus effectively enhance the superconducting diode efficiency. Such properties make 2M-$WS_2$ a great material platform to study the effect of magnetic field on superconducting diode performance[26,27]. Third, the clean van der Waals gap between the two 2M-$WS_2$ nanoflakes can naturally serve as the tunnel barrier for Cooper pairs[28-30], simplifying the fabrication process of high-quality junction devices for investigating the superconducting diode effect (Fig. 1c,d).

To confirm that a superconducting Josephson junction is indeed formed in 2M-$WS_2$ homojunction, we performed the four-terminal electrical current ($I$)-voltage ($V$) measurement, where the voltage $V$ is measured across the superconducting junction between electrodes 2 and 3, as $I_{DS}$ is applied between electrodes S and D (Fig. 1d). One can see in Fig. 1e that the junction shows perfect superconducting state when the electrical current is less than ±30 μA (blue curve), indicating that a Josephson junction with zero tunneling resistance is formed due to the direct Cooper-pair tunneling between two superconducting 2M-$WS_2$ layers. Up to 6 cusp peaks (three pairs) can be clearly observed in the first-order derivative d$V$/d$I$ (red curve) of the original $I$-$V$ data: 1) The smallest $I_c$ values (red triangle pairs, marked as $I_c^+$ and $I_c^-$) correspond to the critical Josephson current of the junction; 2) The two larger $I_c$ values (orange triangle pairs, marked as $I_{c2}^+$ and $I_{c2}^-$) correspond to the critical current of the thin $WS_2$ nanoflake at the bottom (within the area circled by red dashed line in Fig. 1d); 3) The largest two $I_c$ values (blue triangle pairs, marked as $I_{c1}^+$ and $I_{c1}^-$) correspond to the critical current of the thick $WS_2$ nanoflake on the top (within the area circled by yellow dashed line in Fig. 1d). Figure 1g shows a phase diagram of the critical current values as a function of temperature to understand the $I$-$V$ characteristics evolution. One can see that all the $I_c$ values decrease monotonically as the temperature increases, based on which the optimized electrical current range can be selected for further study on the superconducting diode effect.

To investigate the nonreciprocal transport behavior and superconducting diode performance, we carried out controlled experiments in three types of Josephson junctions with different stacking geometries, $WS_2/WS_2$, $WS_2/NbSe_2$ and $NbSe_2/NbSe_2$ (Fig. 2) as comparison. Figure 2a-c shows the stacking geometry, the $I$-$V$ characteristics of the $WS_2/WS_2$ homojunction with different nanoflake thickness, and its half-wave rectification of the supercurrent at 2 K. One can see the nonreciprocal transport behavior with large asymmetric loop in the $I$-$V$ characteristics (Fig. 2b and its inset) , and the value of critical Josephson current $I_c^+$ (flowing from the thin to the thick 2M-$WS_2$ nanoflake, details in Supplementary Fig. 5) is larger than that of $|I_c^-|$ (flowing from the thick to the thin 2M-$WS_2$ nanoflake), indicating the emergence of the superconducting diode effect. Excellent rectification effect with an on-off ratio being ~$10^5$ can be observed based on square wave rectification measurements (Fig. 2c), and the diode performance is stable even after the repeated aging test with >1,000 cycles (Supplementary Fig. 6). Note that all above superconducting diode phenomena were achieved without applying any magnetic field, which behaves as a typical field-free superconducting diode. Unlike previously reported field-free superconducting diodes that need specially designed tunnel barrier that breaks the time-reversal symmetry[6,7,10,11,31], our observed field-free superconducting diode of $WS_2$ Josephson junction does not require this condition. Therefore, the RCSJ model, in which four key factors including the geometric asymmetry, non-equilibrium state, nonlinear quantum capacitance and electronic circuit noise are required for realizing the superconducting diode effect without breaking the time-reversal symmetry, can serve as the scenario explaining our field-free superconducting diode phenomena (more discussion given below). Also, it should be addressed that a finite dissipation is required for such an RCSJ model with nonlinear quantum capacitance to induce nonreciprocal $I_c$ that keeps the time reversal symmetry.

Figure 2d-f shows the device geometry, the *I-V* characteristics, and the half-wave rectification behavior of superconducting $WS_2/NbSe_2$ heterojunction. The reason why we chose 2H-$NbSe_2$ to replace the top thick 2M-$WS_2$ nanoflake is that this 2H-$NbSe_2$ superconductor has the similar $T_c$ and in-plane coherence length, as well as the similar *s*-wave pairing nature[32,33] to those of thick 2M-$WS_2$ superconductor[20-23], so that the superconducting properties of the two materials are compatible and appropriate to form a similar Josephson junction as a comparison. According to the *I-V* characteristics of the $WS_2/NbSe_2$ heterojunction (Fig. 2e), the $I_c^+$ value (from bottom thin 2M-$WS_2$ to top 2H-$NbSe_2$) is still larger than that of $|I_c^-|$ (from top $NbSe_2$ to bottom thin 2M-$WS_2$ nanoflake), indicating the nonreciprocity therein. Similar to the $WS_2/WS_2$ homojunction, the $WS_2/NbSe_2$ heterojunction also demonstrates the field-free superconducting diode behavior, which can serve as the experimental

evidence to confirm the above-mentioned symmetry arguments, where inversion symmetry is broken while time-reversal symmetry remains intact in both diode systems.

As a controlled experiment for comparison, we further performed *I-V* measurements on designed asymmetric $NbSe_2/NbSe_2$ homojunction with different nanoflake thickness (Fig. 2g,h). The values of $I_c^+$ and $|I_c^-|$ in *I-V* characteristics of such a homojunction are almost the same (Fig. 2h), indicating that there is no diode effect without magnetic field, showing remarkably distinct behavior to that in $WS_2/WS_2$ homojunction and $WS_2/NbSe_2$ heterojunction. In sharp contrast, the superconducting diode effect can be generated in such a $NbSe_2/NbSe_2$ homojunction when the external magnetic field is applied to break the time-reversal symmetry, as demonstrated in Supplementary Fig. 7 or other reports[30,34]. Although recent observation field-free superconducting diode effect was also observed in $NbSe_2$ homojunctions[35], such observations might originate from the different defect types and defect concentrations in these $NbSe_2$ samples. But in most cases, the zero-field superconducting diode effect is absent in $NbSe_2$ homojunctions. Since the requested factors of geometric asymmetry, non-equilibrium state and electronic circuit noise have been satisfied in such $NbSe_2/NbSe_2$ homojunction for realizing the field-free diode effect via the RCSJ model yet only the factor of nonlinear quantum capacitance is missing, one can deduce that $NbSe_2/NbSe_2$ homojunction may not have a large enough nonlinear quantum capacitance, probably because the electronic properties of the top thick and bottom thin $NbSe_2$ layers are quite similar[32] and the charge accumulation is almost symmetric with respect to the applied voltage across the junction.

By comparing the zero-field *I-V* characteristics of the above-mentioned three structures, a plausible explanation is that the topological surface state of 2M-$WS_2$ nanoflakes[20-22,24,25] might play a key role on inducing the asymmetric charge accumulation in the Josephson junction even under zero magnetic field, while such an effect is negligible due to the topologically trivial properties of 2H-$NbSe_2$ nanoflakes. Also, note that hysteresis loops caused by linear capacitance effect appear in the *I-V* characteristics of pristine 2M-$WS_2$ nanoflakes (Supplementary Fig. 4a), indicating that the existence of intrinsic Josephson coupling effect in the sample. Correspondingly, one can see that the voltage jumps of $WS_2$/$WS_2$ homojunction in Fig. 2b (~15 mV) is significantly larger than the theoretically-predicted voltage jump of a Josephson junction based on the superconducting gap value of 2M-$WS_2$ nanoflake (~3 mV)[21]. Such results indicate that the Josephson coupling between the neighboring 2M-$WS_2$ layers also exist in the homojunction due to the large van der Waals gap and weak interlayer coupling within 2M-$WS_2$ nanoflakes, and $WS_2$/$WS_2$ homojunction should be considered as the series connection of multiple Josephson junctions. The observations of large voltage jump in *I-V* characteristics can also be observed in other van der Waals superconductors, like Fe(Te,Se) homojunction-based Josephson diodes[31]. Therefore, the comparison among above three device geometries implies that the nonlinear capacitance-coupled RCSJ model might play an important role in realizing the field-free superconducting diode effect.

**Uni-directional polarity of supercurrent**

Normally, in those reported field-free superconducting diode scenario driven by the spontaneous time-reversal symmetry breaking mechanism, the polarity of diode supercurrent will be randomly distributed[36], and could be different when the junction goes to superconducting state during each cooling. While the nonlinear quantum capacitance mechanism in RCSJ model will in contrast induce a fixed polarity of diode supercurrent owing to the fixed inversion asymmetry of the asymmetrically-stacked junction, which provides the smoking-gun evidence to distinguish the nonlinear quantum capacitance mechanism from others. One can see in Fig. 2i that all our measured $WS_2$-based superconducting diodes always host the identical diode polarity with the supercurrent flowing from thin to thick layers with the uni-directional rectification behavior and such a uni-directional rectification is irrespective to the cooling times, which excludes the random nature of supercurrent polarity in spontaneous time-reversal symmetry breaking paradigm and suggests the nonlinear quantum capacitance mechanism with the RCSJ model for 2M-$WS_2$ Josephson junctions (details in Supplementary Figs. 8 to 10 and Supplementary Table 1). Note that, based on the statistical results of supercurrent polarization with >1,000 cycles during the repeated aging test mentioned above (Supplementary Fig. 6), the stable uni-directional nature of these superconducting diodes further confirms the reliability and robustness of our 2M-$WS_2$ superconducting diodes.

Figure 3 shows the evolution of diode efficiency $\eta$ as functions of the temperature, the nanoflake thickness, and the strength of magnetic fields in 2M-$WS_2$-based superconducting diodes. One can see in Fig. 3a that the maximum on-off ratio $\left|\frac{V^-}{V^+}\right|$ for $WS_2$/$NbSe_2$ heterojunction decreases nearly an order of magnitude as temperature increases from 1.6 to 4.2 K. The calculated $\eta$ value ($\eta = \frac{I_c^+ - |I_c^-|}{I_c^+ + |I_c^-|} \times 100\%$) from Fig. 3a also shows the similar decreasing trend as temperature increases. The similar behavior can be observed for $WS_2$/$WS_2$ homojunctions (Fig. 3b and Supplementary Fig. 11). Importantly, the superconducting diode efficiency can be effectively modulated and reach a maximum $\eta$ value up to ~67% at 1.6 K simply by varying the thickness of 2M-$WS_2$ nanoflake layers in such Josephson junctions (Fig. 3c). According to our theoretical simulations of diode performance with different electronic circuit noise level in Fig. 1b, the large fluctuation will inhibit the superconducting diode effect. Therefore, the experimental observations on temperature evolution of $\left|\frac{V^-}{V^+}\right|$ and $\eta$ imply that thermal fluctuation at higher temperature might induce larger electronic circuit noise that disrupts the phase coherence of Cooper pairs, thus weakening the asymmetric supercurrent flow and nonreciprocal charge transport characteristics in 2M-$WS_2$-based superconducting diodes.

To further confirm that the supercurrent polarity in $WS_2$/$WS_2$ superconducting diode is induced by geometric asymmetry of the homojunction, we also presented two controlled experiments: 1) By fabricating a 2M-$WS_2$ homojunction with an inverted geometry (the thick 2M-$WS_2$ nanoflake is placed on the bottom while the thin 2M-$WS_2$ nanoflake on top), we found that the supercurrent polarity is indeed inverted in such a structure (Supplementary Fig. 12). 2) By fabricating another 2M-$WS_2$ homojunction where the thickness of top and bottom 2M-$WS_2$ nanoflakes are almost the same, we found that the superconducting diode efficiency is almost zero in such a geometrically symmetrized device (Supplementary Fig. 13). Therefore, one can see that the direction of supercurrent polarity for our 2M-$WS_2$ superconducting diode is strongly dependent on the geometrical asymmetry during the fabrication process, which is consistent well with the theoretical expectation based on the RCSJ model. Importantly, the large superconducting diode efficiency and large on-off ratio cannot be simultaneously realized for such 2M-$WS_2$-based superconducting diodes. Based on our controlled experiments with up to 20 $WS_2$/$WS_2$ superconducting diode devices (Supplementary Figs. 14, 15), we found that the superconducting diode efficiency can show an increasing trend by increasing the thickness difference between the bottom and top 2M-$WS_2$ nanoflakes, indicating that the larger geometrical asymmetry caused by larger thickness difference between bottom and top 2M-$WS_2$ nanoflakes can induce the larger $\eta$ value of superconducting diode (details in Supplementary Fig. 16). In comparison, the on-off ratio will decrease correspondingly. Such an observation is consistent with the simulation results for nonlinear capacitance-coupled RCSJ model that the large on-off ratio and large superconducting diode efficiency cannot be simultaneously obtained due to the effect of circuit noise (Fig. 1b).

To understand the influence of magnetic field on superconducting diode efficiency, we performed the DC Josephson effect and magneto-chiral anisotropy measurements with in-plane and out-of-plane magnetic fields. By applying an in-plane magnetic field, the DC Josephson effect is expected due to the phase coherence of the superconducting order parameters of the two nanoflakes[37,38] and induces the Fraunhofer pattern in the $I_c$-$\mu_0 H_\parallel$ relation (Fig. 3d). Since the estimated effective thickness $d_{\mathrm{eff}}$ of the 2M-$WS_2$ homojunction is approximately 29 nm based on the equation $I_c(\Phi) = I_c(0)\left|\frac{\sin\left(\frac{\pi\Phi}{\Phi_0}\right)}{\left(\frac{\pi\Phi}{\Phi_0}\right)}\right|$ ($\Phi_0 = h/2e$ is the quantized magnetic flux in superconductors and $\Phi$ is total magnetic flux penetrating through the junction area). Note that not only the Josephson coupling between $WS_2$ nanoflakes, but the Josephson coupling within the $WS_2$ nanoflake will induce the DC Josephson effect, so the not well-developed Fraunhofer pattern in Fig. 3d is not a flaw but a signature of the coexisting intra- and inter-flake Josephson couplings, a characteristic feature of 2M-$WS_2$ due to its weak interlayer interaction. Similar patterns have been reported in other weakly coupled systems such as twisted $Bi_2Sr_2CaCu_2O_{8+\delta}$ homojunction[29] and are distinct from the well-defined patterns in strongly coupled $NbSe_2$ junctions[30]. Also, since the thickness of the Josephson junction (~30 nm) is significantly smaller than the penetration depth $\lambda$ of 2M-$WS_2$ (~270 nm)[39], the applied in-plane magnetic field will penetrate through the entire junction area and the value of $d_{\mathrm{eff}}$ is thus determined by the effective thickness of the entire junction instead of the penetration depth. Experimentally, the fitted $d_{\mathrm{eff}}$ value is close to that of entire Josephson junction (details in Supplementary Figs. 17 to 19 and Supplementary Table 1), one can deduce that the applied in-plane magnetic field can penetrate through the whole junction, which is consistent with the type-II superconductivity nature of 2M-$WS_2$. Such observations indicate that the in-plane magnetic field can serve as a technical way to modulate the superconducting diode efficiency. According to our experimental results shown in Fig. 3e, the $\eta$ value can increase by applying a small in-plane magnetic field, but then decrease as the magnetic field further increases, indicating a non-monotonic modulating effect of in-plane magnetic field on $\eta$ value.

On the other hand, by applying an out-of-plane magnetic field, the diode efficiency can also be enhanced since the field-induced magneto-chiral anisotropy can couple to the superconducting order parameter of the Josephson junction and generate nonreciprocal critical current via breaking time-reversal symmetry[3,37]. To understand the $\eta$ evolution mechanism with out-of-plane magnetic field, we performed field-dependent second-order harmonic measurements to study the magneto-chiral anisotropy index $\gamma = \left|\frac{2R_{xx}^{2\omega}}{\mu_0 H_{\perp} R_{xx}^{\omega} I}\right|$, where $R_{xx}^{\omega}$ and $R_{xx}^{2\omega}$ are first- and second-order longitudinal resistance. Note that a large magneto-chiral anisotropy index $\gamma$ of $6.23\times10^{7}$ $A^{-1}T^{-1}$ indicates that out-of-plane magnetic field can effectively tune the superconducting diode performance (Fig. 3f). To compare our results of $\gamma$ with other systems, we further normalized the $\gamma$ value based on the area (denoted as $A$) of the junction within the flake plane (normalized $\gamma$ is denoted as $\gamma' = \gamma \times A$). The obtained $\gamma'$ = $8.1\times10^{-4}$ $m^2A^{-1}T^{-1}$ within our 2M-$WS_2$ homojunctions is larger than those of most superconducting diode systems[2,3,6,38,40-44] (details in Supplementary Fig. 20 and Supplementary Table 2), indicating the powerful tunability of diode performance in our case. The observed large magneto-chiral anisotropy value of 2M-$WS_2$ homojunction might originate from the asymmetric vortex matching effect that can induce asymmetric vortex lattice motion under external magnetic field[45-47]. Figure 3g summarizes the working temperatures and diode efficiencies of recently reported superconducting diodes[2,5,7,8,10-13,48-60]. One can see that the efficiency $\eta$ of the field-free diode effect of 2M-$WS_2$ homojunctions can reach as high as 67% at 1.6 K, which is comparable to those best record values in twisted trilayer graphene and Pt/V/EuS heterostructures (details in Supplementary Table 3, 4)[7,48].

### 2.4 GHz flip-flop device operation

To explore the device applications of our field-free 2M-$WS_2$ superconducting diodes, we demonstrated the edge-triggered flip-flop devices and their operations with ultrahigh working frequency up to 2.4 GHz (Fig. 4). Figure 4a,b shows the working mechanism of such a flip-flop device: when the amplitude of input sinusoidal signal $I_{in}$ is between $I_c^+$ and $|I_c^-|$, the output voltage of the device is almost zero during the positive half cycle, and the junction maintains the zero-resistance superconducting state (light blue area, marked as "S"); however, for the negative part of $I_{in}$, the junction turns to the metallic state when the amplitude of $I_{in}$ is larger than $|I_c^-|$ of the Josephson junction (light orange area, marked as "M"), thereby generating a sharp pulse signal $V_{out}$. Therefore, by tuning the amplitude of input sine signal and the $\eta$ value of the superconducting diode, the pulse width and duty cycle of the output signal can both be efficiently modulated, demonstrating an edge-triggered flip-flop functionality. The pulse width and duty cycle within the flip-flop device can also be designed by changing the input voltage and working temperature, as well as external magnetic fields, as shown in Supplementary Figs. 21 to 23.

Figure 4c shows the pulse signal generation results based on our flip-flop device, in which the pulse signal with a cusp peak (inset of Fig. 4c) can be obtained owing to the above-mentioned rectification mechanism. Note that the generated pulse signal is highly repeatable with the same pulse width and duty cycle during the consecutive measurements over 2500 s. In sharp contrast to the rectification results of the Josephson junction, the top and bottom superconducting layers can only show reciprocal behavior, in which 2H-$NbSe_2$ nanoflake remains in its zero-resistance state and 2M-$WS_2$ nanoflake becomes metallic state for both electrical current flow directions (Fig. 4d). Figure 4e shows the output pulse generation behavior in our 2M-$WS_2$-based flip-flop device under different types input periodic signals. One can see that three types of input signals (sine, triangular and chainsaw wave) can induce a similar pulse feature in the corresponding output signals, indicating that such a superconducting diode-based pulse generator can have a large input tolerance, while the output signal under square-wave excitation behaves different from other excitation waveforms due to the longer dwell time at minimum voltages and more higher-order components of the square wave in Fourier transform compared to the other three types of input waves. Note that the generated pulse signals are always along the negative direction due to the uni-directional supercurrent polarity, which is consistent with the fixed polarity predicted by the RCSJ model mentioned above (Fig. 2i). For $WS_2$-based superconducting diodes, according to the magnitude estimation of $I_c R \sim \frac{\Delta}{e} \sim 1$ mV, one can calculate the time scale of voltage oscillation (in a voltage state of the non-equilibrium RCSJ model) is $\frac{2eI_c R}{\hbar} \sim 10^{-12}$ s, corresponding to 1 THz. Therefore, a rough theoretical estimation of the upper limit of the rectification frequency based on our 2M-$WS_2$ superconducting diode device with the underlying RCSJ mechanism is an order of magnitude smaller than the time scale of voltage oscillation, i.e., 100 GHz.

Therefore, one can expect the upper limit of the working frequency can reach sub-THz (details in Methods section and Supplementary Figs. 24, 25 and Supplementary Table 5).

Figure 4f shows the rectification results of the input 2.4 GHz square wave together with their Fourier transform results. One can see that the duty cycle of the output signal is close to 50% (same as that of the input signal), yet the on-off ratio at such a high frequency shows a decreasing trend compared to the low frequency situations (Fig. 4c-e). On one hand, at all measured frequencies, a cusp pulse signal can always be generated based on the input sine wave (Supplementary Figs. 26 to 32 and Supplementary Table 6), indicating that our designed flip-flop device demonstrates a broadband working frequency range with 12 orders of magnitude from $10^{-3}$ Hz (Supplementary Fig. 32) to $10^{9}$ Hz (Fig. 4f). On the other hand, even up to the high frequency of 2.4 GHz, the output voltage of the flip-flop device can perform similar behavior for sine and triangular input waves (Supplementary Fig. 31), demonstrating the input tolerance of our flip-flop device can maintain up to the high frequency regime. Figure 4g summarizes the highest working frequencies of our and other superconducting diode devices (details in Supplementary Table 5)[5,10,49,54,61-64], where our flip-flop design operates at a frequency nearly two orders of magnitude greater than the previous state-of-the-art. This implementation of $WS_2$-based superconducting diodes in a flip-flop device paves the way for future high-frequency and broadband applications in superconducting electronics.

## Conclusions

In summary, we demonstrated field-free superconducting diodes and high-frequency edge-triggered flip-flop devices with high-frequency pulse signals up to 2.4 GHz and a broadband working frequency of 12 orders of magnitude. Unlike the conventional approach of the field-free superconducting diode effect that requires the breaking of the time-reversal symmetry, our work demonstrates that geometric asymmetry in non-equilibrium Josephson junctions can induce field-free rectification effect described by the RCSJ model that keeps the time-reversal symmetry, providing an alternative design principle of high-frequency superconducting diodes and device applications. The application of superconducting diodes transcends simple rectification by leveraging their unique quantum properties—zero resistance, Josephson effects, and nonreciprocal quantum transport—to solve critical challenges in superconducting logic circuits, quantum computing, ultrahigh-frequency telecommunication, and the creation of low-power, nonlinear elements for brain-inspired superconducting neuromorphic computing.

## Methods

### Simulations on the nonlinear quantum capacitance-induced Josephson diode effect

Our simulation for field-free superconducting diode effect is based on the RCSJ model with noise current and the nonlinear quantum capacitance caused by inversion symmetry breaking[18]. In the absence of inversion symmetry, the accumulated charge $Q$ on the junction can be different under opposite voltage $V$ satisfying the relation $V = Q/C + aQ^2 + bQ^3$ up to third order, where $C$ is the classical linear capacity and $a, b$ come from second-order and third-order nonlinear response of $Q$ on $V$, called "quantum capacities". In the RCSJ model, the charge $Q$ and phase flux $\frac{h}{2e}\phi$ are a pair of canonical conjugate variables satisfying the following equations

$$\frac{h}{2e}\dot{\phi} = V = \frac{Q}{C} + aQ^2 + bQ^3,$$

$$\dot{Q} = I_{\mathrm{dc}} - J\sin\phi - \frac{V}{R} + I_{\mathrm{noise}},$$

where $I_{\mathrm{dc}}$ is the external DC current, $R$ is the resistance of the normal junction, $J$ is the critical current without noise, $a$ and $b$ represent the effect of the nonlinear relation between the voltage $V$ and charge $Q$, $I_{\mathrm{noise}}$ is the noise current supposed to be white noise satisfying $\langle I_{\mathrm{noise}}(t) I_{\mathrm{noise}}(t')\rangle = D\delta(t - t')$. By numerical integration, we obtain the time-dependent voltage $V(t)$. Its time average gives the dc voltage $V_{\mathrm{dc}} = \langle V(t)\rangle$. Due to the asymmetric charge accumulation, the re-trapping currents $I_{\mathrm{r}\pm}$ become nonreciprocal, although the critical currents $I^0_{\mathrm{c}\pm}$ (without noise) keep reciprocal. For the currents $I_{\mathrm{r}} < I < I^0_{\mathrm{c}}$, there are two metastable states: one with zero voltage, and one with finite voltage. Weak noise will mix these two states randomly and lead to nonreciprocal experimental critical currents $I^{\mathrm{noise}}_{\mathrm{c}+} \neq I^{\mathrm{noise}}_{\mathrm{c}-}$, while strong noise will spoil the two metastable states completely, leading to ill-defined $I_{\mathrm{c}}$. Note the time scale of the variance of $V(t)$ is very short with order of $\frac{h}{2eV} \sim 10^{-12}$ s (for $V \sim$ 1mV, corresponding to ~1 THz), and thus has little effect on the AC measurement with frequency much smaller than 1 THz, otherwise, the Josephson oscillation will be measured instead of the rectification phenomenon. In this mechanism, the field-free diode effect is robust and can be well controlled by junction fabrication without the necessity of time-reversal symmetry breaking. Therefore, it provides a more flexible way to design a Josephson diode in future applications.

**Superconducting diode device fabrication**

2M-$WS_2$ and 2H-$NbSe_2$ nanoflakes were obtained by mechanical exfoliation of layered bulk crystals in a $N_2$ glovebox. The nanoflakes were then stacked onto Si/$SiO_2$ substrates via a dry transfer technique with a polydimethylsiloxane (PDMS) film. Polymethyl methacrylate (PMMA) was then spin-coated on the substrate and put in a vacuum drying oven kept at <0.01 MPa without heating the sample, which can remove the solvent in PMMA but avoid introducing any phase transition in 2M-$WS_2$ nanoflakes. A standard electron beam lithography (EBL) process was used for device fabrication, and 5/60 nm-thick Ti/Au electrodes were evaporated. The developed devices were cleaned by Ar plasma for 30 s before metal evaporation to achieve better electronic contact.

**Low-temperature transport measurements for superconducting diodes**

The electronic transport properties of the devices were measured with a temperature down to 1.5 K and a magnetic field up to 12 T in an Oxford Instruments Teslatron. The four-terminal resistance and *I-V* characteristics of the devices were studied by both the direct current method and the alternating current method. For the direct current method, we used a sourcemeter (Keithley 2400) and a nanovoltmeter (Keithley 2182A) for nonreciprocal *I-V* relations and half-wave rectification measurements, while for the alternating current method, SR830 lock-in amplifiers (Stanford Research

Systems) at frequency $f = 13$ Hz were used for second-harmonic measurements and DC Josephson effect measurements.

**High-frequency measurements of edge-triggered flip-flop device**

We used two setups for measuring the performance of our trigger circuits. We used an SR830 lock-in amplifier to form an alternating current injection, and a Keithley 2182A nanovoltmeter to measure pulse signals at the low frequency limit (~0.002 Hz) to capture the shape of the generated pulse. We further applied a waveform generator (Keysight 33600A Series) to generate programmed sinusoidal input, and an oscilloscope (Tektronix TBS 1202B) to measure the output signals at high frequency limit up to 80 MHz to investigate the performance of the device at high frequency. An SR560 preamplifier (Stanford Research Systems) was also used for frequencies < 1 MHz to amplify the diode's output signal for better analysis of waveform patterns. While an arbitrary waveform generator (Tektronix AWG7122B) and a digital oscilloscope (Tektronix DPO75902SX) was applied for frequencies up to 2.4 GHz to determine the highest working frequency of our device. All setups include a resistor of ~100 Ω in serial connection with the device for protection.

**Data availability**

All data needed to evaluate the conclusions in the paper are presented in the paper or the supplementary materials and are available from the corresponding author upon reasonable request.

**Acknowledgements**

The authors would like to acknowledge the support by the National Natural Science Foundation of China (Nos. 92365203 (H.T.Y., Q.H.W., H.J.Z.), U24A6002 (H.T.Y.), 12374147 (Q.H.W.), 52302180 (J.W.H.), 12274205 (D.W.)), the National Key Research and Development Program of China (Nos. 2024YFA1408104 (H.T.Y., Q.H.W.), 2021YFA1202901 (J.W.H.), and 2022YFA1403201 (D.W.)), the Quantum Science and Technology-National Science and Technology Major Project (No. 2021ZD0302502 (H.T.Y.)), the Natural Science Foundation of Jiangsu Province (Nos. BK20253012 (H.T.Y.), BK20243011 (H.T.Y.), BK20233001 (H.T.Y.)), Shanghai Rising-Star Program (No. 23QA1410700 (Y.Q.F.)), the Science and Technology Commission of Shanghai Municipality (No. 24LZ1401000 (Y.Q.F.)).

**Author Contributions**

H.T.Y. and Q.H.W. conceived the project. X.Y.B., H.Y.L. and A.S.Y. prepared the superconducting diode devices. X.Y.B. and H.Y.L. performed low-temperature electronic transport measurements. A.S.Y. and Q.Z.S. performed AFM measurements. H.Y.L., J.W.H., S.C.Y., Y.C.S., W.J. performed high-frequency measurements. Y.Q.F. and F.Q.H. prepared the high-quality 2M-$WS_2$ crystal. D.W., Q.H.W. and H.J.Z. presented theoretical modeling of superconducting diode effect. X.Y.B., H.Y.L., G.Y.C. and H.T.Y. analyzed the data. X.Y.B., D.W. and H.T.Y. wrote the manuscript with inputs from all authors.

**Competing interests**

The authors declare no competing interests.

**Correspondence and requests for materials** should be addressed to H.T.Y..

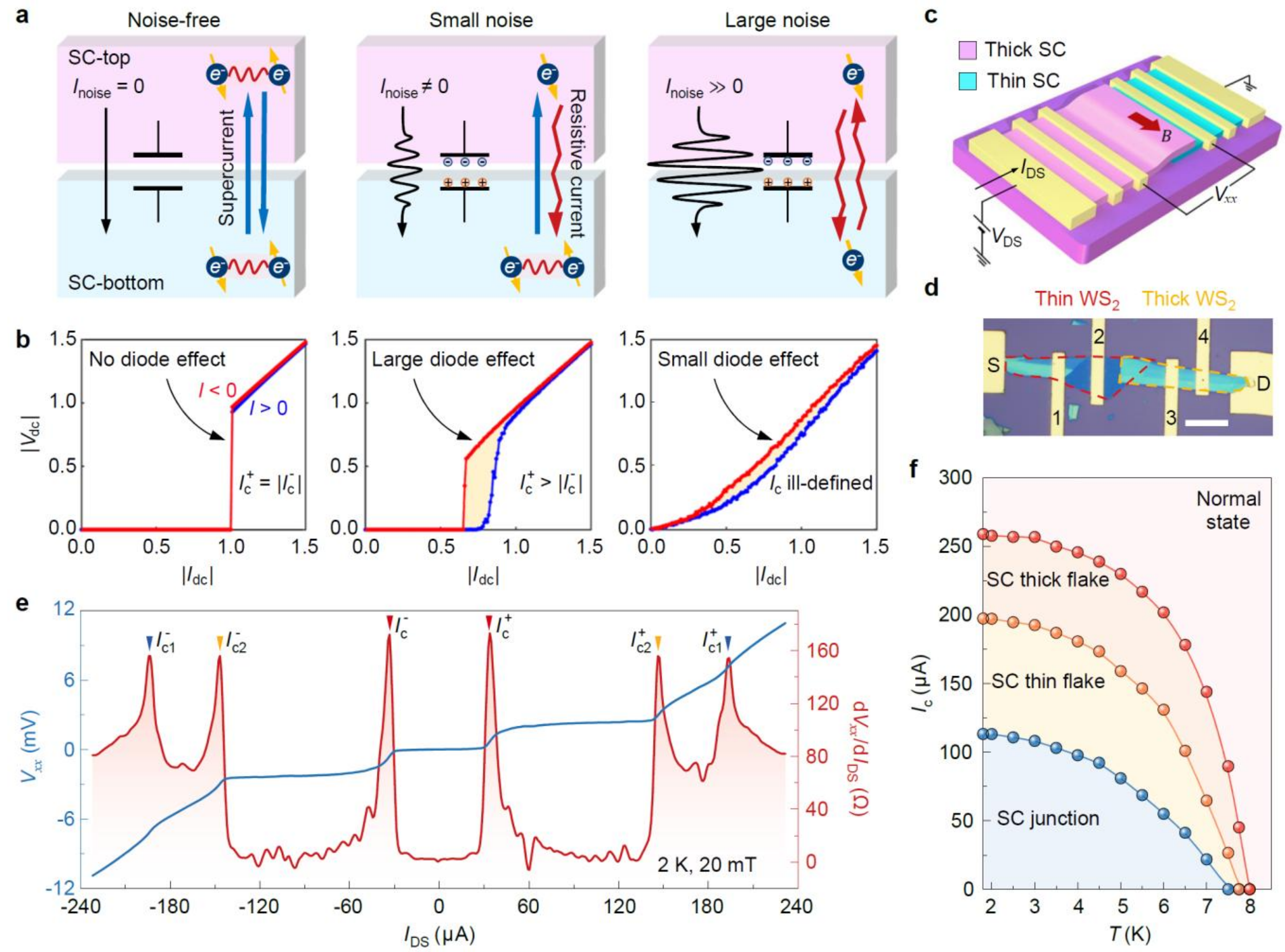

**Fig. 1 | Superconducting diode effect without breaking time-reversal symmetry in 2M-$WS_2$ Josephson junction. a**, Schematic illustration of geometrically asymmetric Josephson junctions under increasing electronic circuit noise $I_{noise}$. **b**, Simulation results based on RCSJ model of the device structures shown in (a), where the field-free superconducting diode effect can be observed under appropriate electronic circuit noise level due to the existence of nonlinear quantum capacitance up to third order. **c**,**d**, Schematic illustration and optical microscopy image of a typical superconducting diode device, where “SC” represents superconductor, electrodes “S” and “D” represent the source and drain of electrical current flow, and electrodes 1 to 4 are voltage probes for measuring the four-terminal resistance of the junction and both nanoflakes. Scale bar is 10 μm. **e**, The *I*-*V* characteristic (blue curve) and its first-order derivative (red curve) of the Josephson junction, where 3 pairs of the critical current

values can be confirmed: $I_c^+$ and $I_c^-$ correspond to the critical supercurrent of the Josephson junction (red triangles); $I_{c1}^+$ ($I_{c2}^+$) and $I_{c1}^-$ ($I_{c2}^-$) correspond to the critical supercurrent of thick (thin) 2M-$WS_2$ nanoflake, indicated by blue (orange) triangles. **f**, The phase diagram demonstrating the critical current and temperature dependence of the Josephson junction based on the *I*-*V* characteristics.

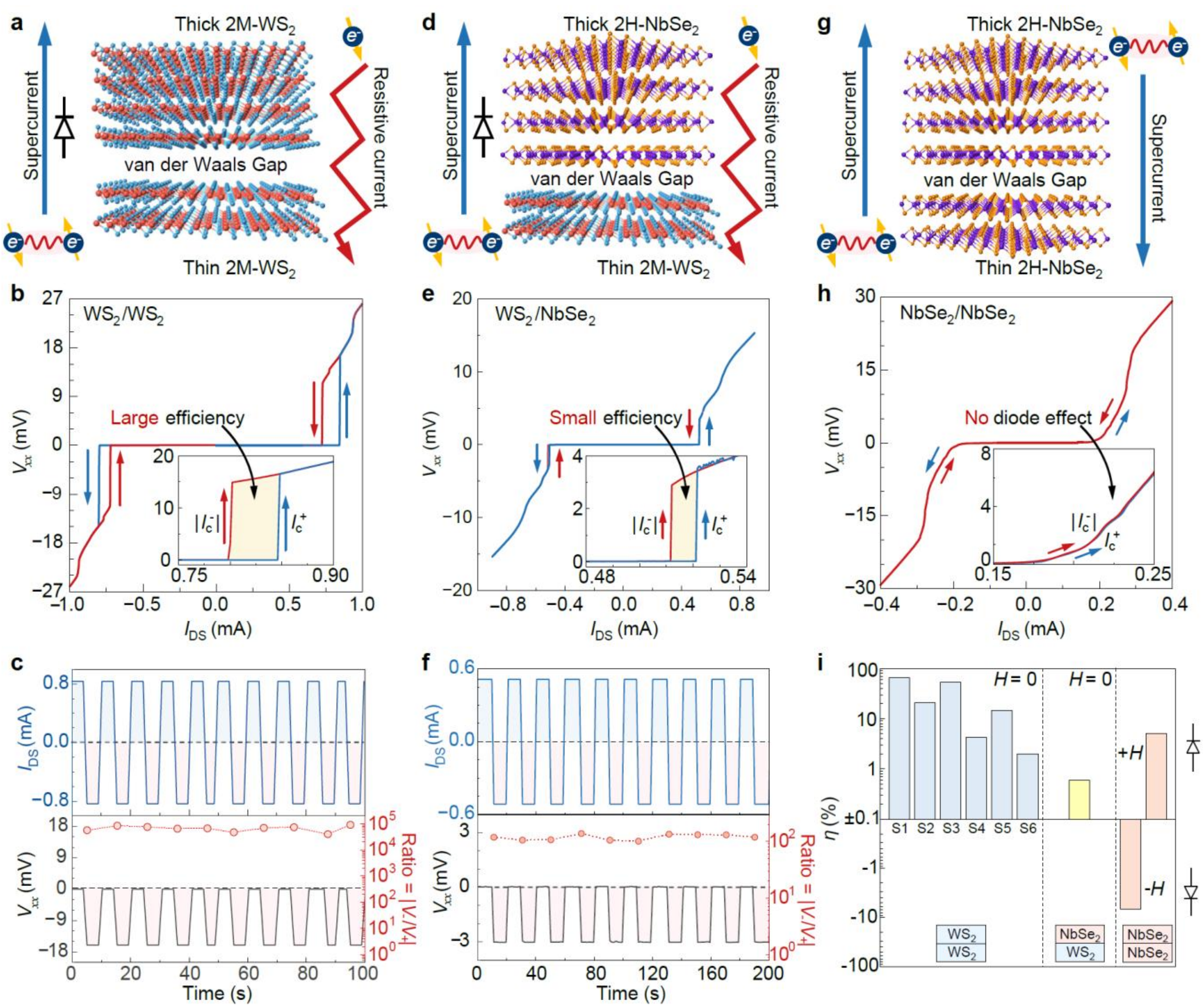

**Fig. 2 | Geometry comparison for superconducting diode effect among three different device configurations. a-c**, Schematic illustration, *I-V* characteristics, and half-wave rectification results of Josephson junction formed by thin (bottom) and thick (top) 2M-$WS_2$ nanoflakes. Large superconducting diode efficiency with on-off ratio up to ~$10^5$ (without applying magnetic field) can be confirmed within such a structure. **d-f**, Schematic illustration, *I-V* characteristics, and half-wave rectification results of Josephson junction formed by thin 2M-$WS_2$ nanoflake (bottom) and thick 2H-$NbSe_2$ nanoflake (top). Smaller superconducting diode efficiency with on-off ratio close to ~$10^2$ (without applying magnetic field) can be confirmed within such a structure. **g**,**h**, Schematic illustration and *I-V* characteristics of Josephson junction formed by thin (bottom) and thick (top) 2H-$NbSe_2$

nanoflakes as a blank control. No field-free superconducting diode effect can be observed in such a structure. **i**, Statistics of the supercurrent polarities of our superconducting diodes. All the measured $WS_2$-based field-free superconducting diodes shows the same supercurrent direction flowing from thick to thin nanoflakes, while the polarity of field-driven $NbSe_2$/$NbSe_2$ superconducting diode depends on the direction of the applied out-of-plane magnetic field.

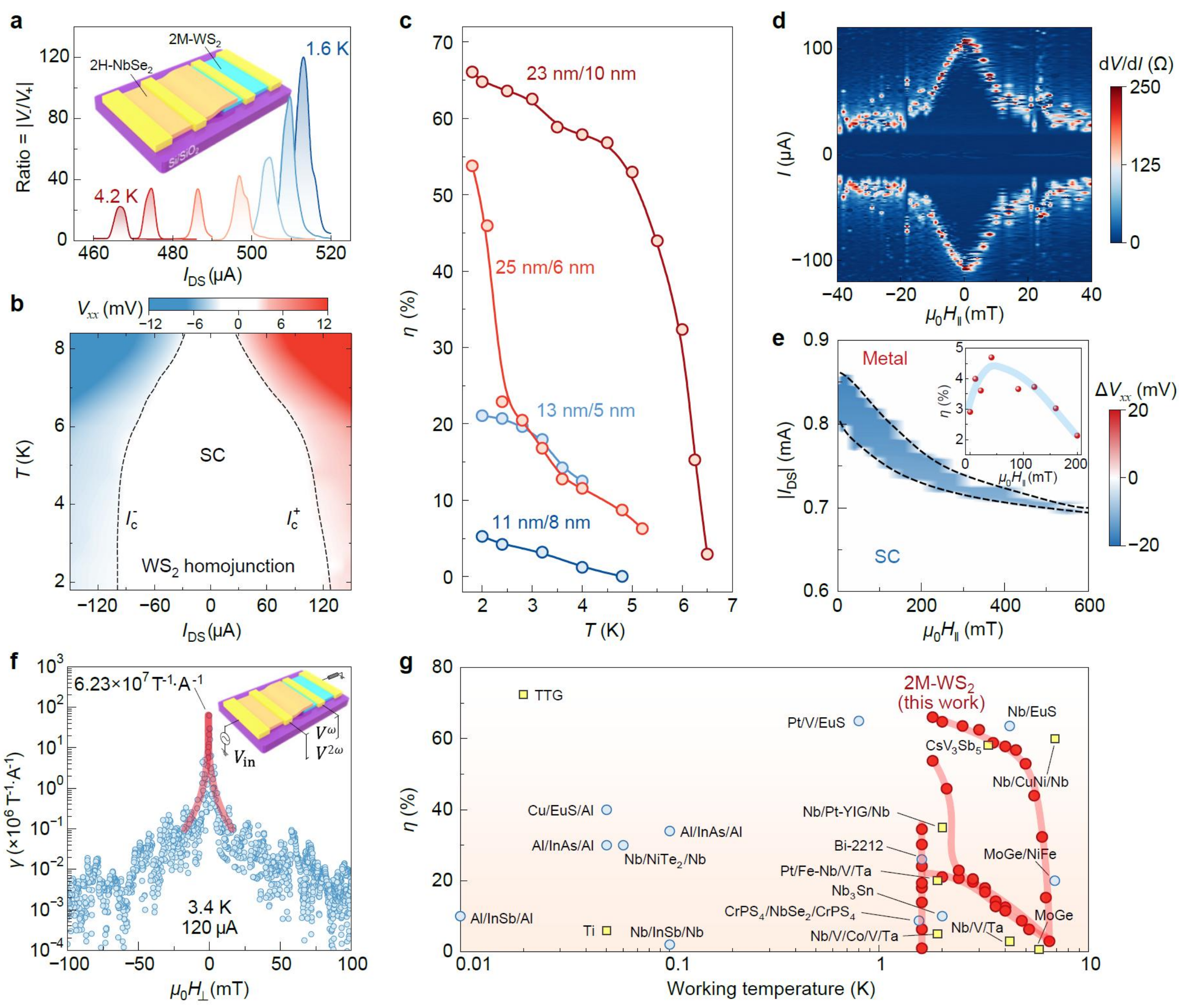

**Fig. 3 | Modulation of superconducting diode effect in 2M-WS2-based Josephson junctions. a**, Temperature-dependent on-off ratio $\left|\frac{V^-}{V^+}\right|$ for a $WS_2/NbSe_2$ superconducting diode. **b**, Temperature dependence of the critical supercurrents $I_c^+$ and $I_c^-$ (dashed lines guide for eyes) for a $WS_2/WS_2$ superconducting diode, and the asymmetric shape indicates the nonreciprocal behavior from the diode effect. **c**, Temperature-dependent diode efficiency $\eta$ of 2M-$WS_2$ homojunctions with different thickness. **d**, The DC Josephson effect and Fraunhofer pattern of critical supercurrent under an in-plane magnetic field for a 2M-$WS_2$ homojunction. **e**, Modulation of the superconducting diode effect under in-plane magnetic field, determined by $\Delta V_{xx} = V_{xx}(+I) - |V_{xx}(-I)|$. Inset: Magnetic field-

dependent diode efficiency $\eta$. **f**, Second-order harmonic measurements for $WS_2$ homojunction and the fitted magneto-chiral anisotropy index $\gamma$ under out-of-plane magnetic field at 3.4 K. Inset: Schematic device geometry for second-order harmonic measurement. **g**, Comparison of temperature-dependent superconducting diode efficiencies of ours and other reported systems. The translucent red curves are trend lines of superconducting diode efficiency evolutions guiding for eyes. Yellow squares (blue circles) represent the field-dependent (field-free) superconducting diode effects, and red circles are from several 2M-$WS_2$ homojunction devices in this study. Other data points are adopted from references[2,5,7,8,10-13,48-60].

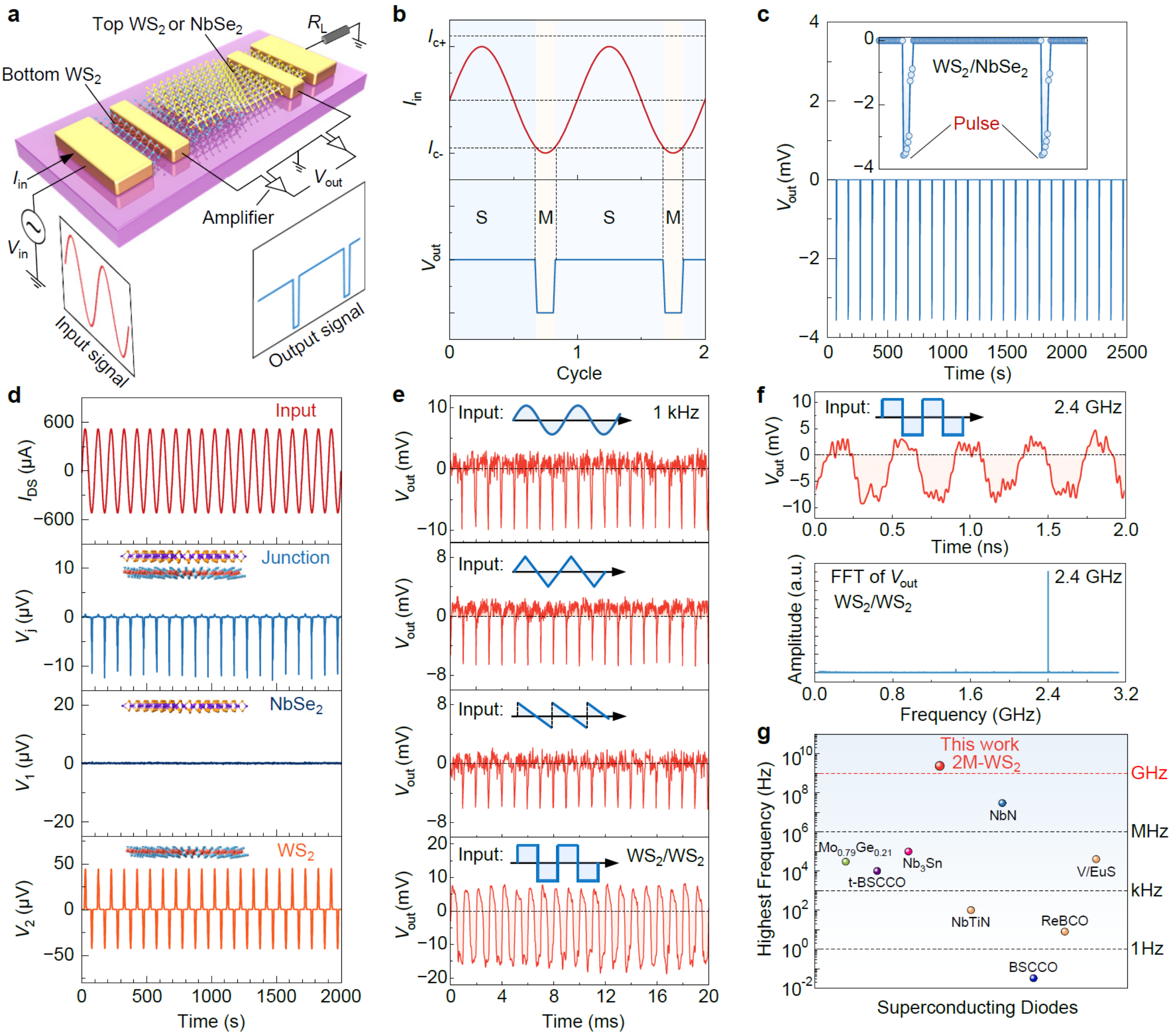

**Fig. 4 | High frequency and broadband edge-triggered flip-flop device from 2 mHz to 2.4 GHz.**

**a**, Schematic illustration of an edge-triggered flip-flop device structure. **b**, Operating mechanism for pulse signal generation within an edge-triggered flip-flop device. When the amplitude of the input electrical current is smaller (larger) than $I_c^+$ ($I_c^-$) value of the superconducting diode, the Josephson junction will maintain the superconducting state (turn to the metallic state), thereby generating pulse signals. **c**, Edge-triggered flip-flop operation of $WS_2/NbSe_2$ superconducting diode at 1.6 K without any external magnetic field. Inset: local enlargement of two neighboring pulses. **d**, Edge-triggered flip-flop operation and pulse signal generation of $WS_2/NbSe_2$ superconducting diode under an external magnetic field of 45 mT, with input signal (red), output signal from the Josephson junction (light blue), voltage between $NbSe_2$ nanoflake (dark blue), and voltage between $WS_2$ nanoflake (orange). **e**, Edge-triggered flip-flop operation and pulse generation under different types of input signals (1 kHz, with the same maximum input voltage) for $WS_2/WS_2$ superconducting diode. **f**, Edge-triggered flip-flop operation of $WS_2/WS_2$ superconducting diode at 2.4 GHz. Upper: Highest output signal under the input square wave signal at 2.4 GHz. Lower: Fast Fourier transform of the output signal $V_{out}$. **g**, Comparison of the highest working frequencies in our flip-flop devices and other reported superconducting diodes[5,10,49,54,61-64].